\documentclass[conf]{IEEEtran}
\usepackage{blindtext, graphicx}
\usepackage{graphicx}
\usepackage{xcolor}
\usepackage{caption}
\usepackage{multirow}
\usepackage{placeins}
\usepackage{subfig}

\usepackage[linesnumbered,ruled,vlined,lined,boxed,commentsnumbered]{algorithm2e}
\include{pythonlisting}
\usepackage{cite}

\usepackage[cmex10]{amsmath}
\usepackage{url}
\usepackage{tabto}
\usepackage{amsmath}

\begin{document}
\bstctlcite{IEEEexample:BSTcontrol}
    \title{Developing Mobility and Traffic Visualization Applications for Connected Vehicles}
  \author{Mohammad~A.~Hoque, Noah~Carter, Md~Salman~Ahmed, Jacob~Hoyos, Matthew~Dale, JT Blevins, Nick Hodge, Nusrat Chowdhury 
\\
Department of Computing\\
   East Tennessee State University\\
   \{carterns, hoquem, ahmedm\}@etsu.edu
}

\author{\IEEEauthorblockN{
Noah~Carter\IEEEauthorrefmark{1},
Md~Salman~Ahmed\IEEEauthorrefmark{2}
Jacob~Hoyos\IEEEauthorrefmark{1},
Matthew~Dale\IEEEauthorrefmark{1},
JT Blevins\IEEEauthorrefmark{1},
Nick Hodge\IEEEauthorrefmark{1},
Nusrat Chowdhury\IEEEauthorrefmark{1},
Mohammad~A.~Hoque\IEEEauthorrefmark{1}}\\
\IEEEauthorblockA{\IEEEauthorrefmark{1}Department of Computing, East Tennessee State University\\
\IEEEauthorrefmark{2}Department of Computer Science, Virginia Polytechnic Institute and State University\\
\IEEEauthorrefmark{1}\{carterns, hoyosj, dalems, blevinsjt, hodgen, chowdhury, hoquem\}@etsu.edu \\\IEEEauthorrefmark{2}ahmedms@vt.edu}}

\maketitle

\begin{abstract}
This technical report is a catalog of two applications that have been enhanced and developed to augment vehicular networking research. The first application is already described in our previous work \cite{Noah2018SoutheastCon}, while the second one is a desktop application that was developed as a publicly-hosted web app which allows any Internet-connected device to remotely monitor a roadway intersection's state over HTTP. This collaborative work was completed under and for the utility of ETSU's Vehicular Networking Lab. It can serve as a basis for further development in the field of connected and autonomous vehicles. 
\end{abstract}

\section{Application--1: Multi-hop Connectivity Simulation}

\IEEEpeerreviewmaketitle


\subsection{Introduction}

The Basic Safety Message (BSM) is a standardized communication packet that is sent every tenth of a second between connected vehicles during vehicle-to-vehicle (V2V) communication. Indeed, one of the primary requirements for all vehicle safety applications is the mandatory broadcast of Basic Safety Messages (BSMs) at this time interval. BSMs contain a wide range of data describing the sending vehicle's state in time, such as speed, location, and the status of the turn signal \cite{safety}. When vehicles and human/non-human drivers have more information about the driving environment around them (such as is provided by BSMs), it follows that they can make safer decisions on the road and drive with higher efficiency. In general, vehicle-to-vehicle (V2V)  and vehicle-to-infrastructure communications like BSM have major implications for improving automotive transportation.

During studies of vehicle connectivity, BSMs can be collected and aggregated into datasets. Currently, many BSM datasets are available from the connected vehicle testbeds of the U.S. Department of Transportation. (A rudimentary analysis of one such dataset can be found near the end of this paper.) However, without a proper visualization tool, it can be very difficult to quickly and visually analyze the spatio-temporal distribution of the vehicles. To fill this need, a web application has been collaboratively developed (by Salman Ahmed and Noah Carter) which can ingest a raw BSM dataset and display a time-based simulation of vehicle movement. The simulation has also been designed to display the ad-hoc network connectivity of the vehicles, each of which is assumed to utilize multi-hop communication over Dedicated Short Range Communication (see below).

Dedicated Short Range Communication (DSRC) is a mode of communication which can, over the distance of approximately 1000 meters [citation needed], be utilized for a variety of safety-critical applications in, for instance, vehicle-to-vehicle (V2V) and vehicle-to-infrastructure (V2I) communication. (Note that such applications follow standard communication protocols defined by IEEE 1609.x. and the new SAE J2735.) In brief, DSRC is used to send BSMs. However, because the range of DSRC is in many cases less than 1000 meters, in order to extend the range it is necessary and useful to employ `multi-hop' communication. In multi-hop communication, messages traverse an ad hoc network of vehicles (that act as message relays) in order to reach their destinations. Thus, rather than going directly from the source vehicle to the targets, the messages `hop' along an ad hoc network of closer vehicles to eventually reach as many targets as possible.

This section of the technical report provides details about the simulation application, including an explanation of the multi-hop partitioning algorithm used to classify the vehicles into separate ad hoc network partitions. In addition, a performance analysis for the simulation is included, in which it is suggested that calculating a complete connectivity matrix with the multi-hop partitioning algorithm is computationally expensive when the number of vehicles is high.

\subsection{Related Works}

Over the past two decades, researchers in the wireless communication and networking community have been focusing on the technologies for vehicular communication. Researchers of East Tennessee State University's Vehicular Network Lab has been working on several layers of the wireless network all the way up to the application layer of vehicular network \cite{hoque2009multiple,khaled2015train,hong2010exploring,hoque2012biostar,hoque2010delay,hoque2008call,hoque2011channel,hoque2012innovative,hoque2012protocol,hoque2009interference}.

Below we mention some of the work related to the two applications mentioned in this report:

\subsubsection{BSM Data Analysis}

As stated above, some BSM databases are publicly available and are interesting subjects for data analysis. Our previous research has utilized BSM data for analyzing mobility patterns and developing various safety-critical as well as mobility applications in connected vehicle environments \cite{hoque2012analysis,elbery2015integrated,elbery2015vnetintsim,ahmed2016comparative,ahmed2016demo,ahmed2016partitioning,ahmed2017demo,jordan2017poster,Ahmed,hoque2018impact,ahmed2018cooperative,hoque2017safari,hoque2013methods}. This includes, for example, the Safety Pilot Model Deployment, in which tens of Michigan volunteers' vehicles were mounted with BSM-broadcasting devices. (See the final section of this technical report for a basic analysis of that particular dataset.) One aspect of such data analysis is visualization and simulation. Allowing an analyst to see the data in a helpful and intuitive format enables the analyst to make more informed and strategic decisions \cite{Safety, Liu}.

The application described in this report is intended to assist an analyst in visualizing location-related aspects of BSM data, specifically vehicle locations and vehicle connectivity partitions. To display the latter, it is necessary to calculate the partitions using a multi-hop partitioning algorithm.

\subsubsection{Multi-Hop Partitioning Algorithm}

Hoque et al. \cite{hoque} introduce a multi-hop partitioning algorithm in their 2013 article. (Note that the algorithm used in this report is similar but does not exactly match the earlier version.) The purpose of this algorithm is to form a matrix that distinguishes each ad hoc partition. That is, the algorithm results in a square matrix which has grouped the vehicles based on whether or not they are within multi-hop communication range of each other.

Again, for clarity: under multi-hop communication vehicles can communicate directly, but they may also do so indirectly via one or more intermediary vehicles through which data and messages can pass (or `hop'). In the output of the algorithm, all vehicles that can communicate with each other by one of these two means are grouped together as a partition. The result is a collection of partitions.  \cite{hoque}

\begin{figure}[h!]
\centering
\includegraphics{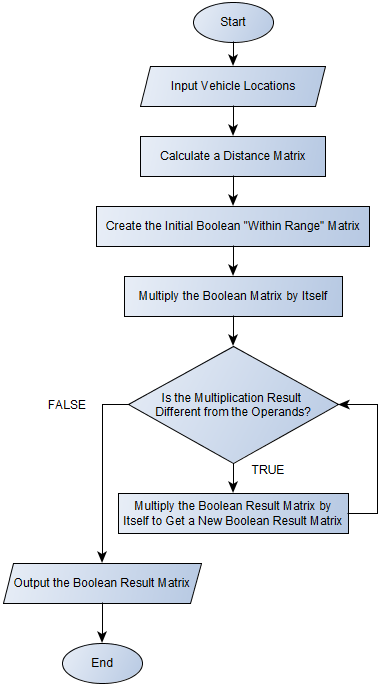}
\caption{Flowchart for a Multi-Hop Partitioning Algorithm, Simplified from Hoque et al.'s \cite{hoque} Implementation.}
\label{multihopFlowchart}
\end{figure}

The algorithm begins (see Figure \ref{multihopFlowchart}) by reading in vehicle locations and calculating a square, symmetric `distance matrix.' Each element $a_{i,j}$ of the distance matrix is the calculated distance between the two vehicles that are represented by row $i$ and column $j$. Each entry of the distance matrix is then translated into a boolean value by determining whether it is less than or greater than the predetermined DSRC range (i.e., the maximum distance a given DSRC device can broadcast communications, which defaults to 1000 meters). Those distances that are less than the range become $1$; others become $0$. (This boolean matrix is the first `connectivity matrix;' it describes whether or not vehicles can communicate.) \textbf{Using boolean algebra,} the newly-formed boolean matrix is then multiplied by a copy of itself. This multiplication represents the first `hop,' and the result is a second `connectivity matrix.'

The algorithm then proceeds with a series of boolean matrix multiplications, each of which represents a new hop. Every iteration has the possibility of stringing together and consolidating more vehicles into a partition. Once a matrix multiplication results in a connectivity matrix identical to the operands, then the connectivity matrix is considered finalized. Any two vehicles $a_i$ and $a_j$ which have a $1$ at $a_{i,j}$ are said to be connected together by multi-hop communication. They share the same connectivity partition.

\subsection{The Simulation}
\subsubsection{Concept and Purpose}

The purpose of this section's application is to ingest BSM data and use it to display a simulation of vehicle activity and connectivity with respect to time. The simulator takes as input Basic Safety Messages from multiple vehicles. Ideally, these are BSMs from vehicles that were concurrently producing BSM output and that therefore have similar timestamps. Once these BSMs are uploaded as a CSV, for any given timestamp the simulator can display the position of each vehicle as a pin on a map (using the Google Maps API). The user can iteratively progress from one timestamp to the next, watching the pins move along the roads on the map (see Figure \ref{low}).

\begin{figure}[h!]
\centering
\includegraphics[width=3in]{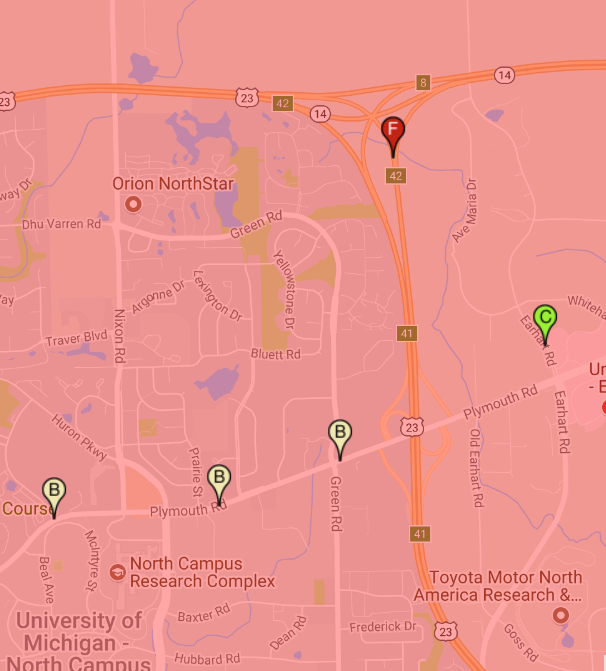}
\caption{Vehicle locations represented using markers of different colors}
\label{pins}
\end{figure}

In addition to its position, each pin has a color and a character(s). Pins with matching color-character combinations are recognized to be in the same connectivity partition--they are able to send data to each other via multi-hop communication (see Figure \ref{pins}). For example, if two pins are within the DSRC range of each other, they can communicate with a single hop; they will therefore share the same color-character combination for each timestamp in which they remain within range. If two pins are not within the DSRC range, then they cannot communicate unless there exists a middle pin that they can both reach via DSRC. If there is at least one mutually-reachable middle pin between the two pins, the external pins can use the middle pin to help broadcast their message, creating an ad-hoc network and successfully utilizing multi-hop communication. Obviously, the multi-hop algorithm is used by the simulation to determine the connectivity matrix and thus the color-character combinations.

In Fig. \ref{pins}, vehicle locations at the current timestamp (as stored in the ingested BSMs) are represented with pins. The three vehicles at the bottom left share the same connectivity partition; they have the same color and character because they can reach each other via multi-hop communication.With the passing of each timeframe, the colors and characters on the pins will change. This is because vehicles move in and out of range, alternately leaving and mingling among connectivity partitions. In the simulation, the range of DSRC devices--which determines the maximum distance of a single hop--can be adjusted by the user; the default DSRC range, which is based on prior research, is 1000 meters. However, in cities it is expected that the range could decrease.

Note: the simulation was created with C\# and ASP.NET; Javascript calls were made to the Google Maps API.

\subsubsection{Usage}

\begin{figure}[h!]
\centering
\includegraphics[width=3in]{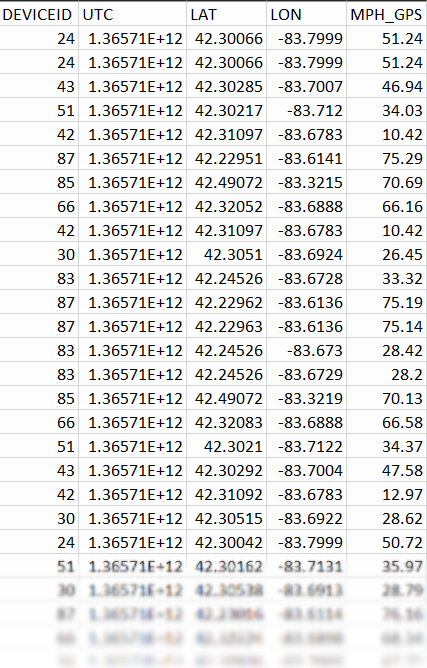}
\caption{Expected CSV Input Format for Simulation}
\label{expected_input_format}
\end{figure}

An example of the expected input format is a large CSV with the structure of Figure \ref{expected_input_format}. Each row should be an abbreviated, cleaned form of a traditional BSM. All BSM fields are removed by the analyst beforehand except the remaining five listed. After uploading, the user must only click the ``Run Simulation" button to proceed automatically or the ``Next Step'' button to iteratively step through the timestamps one-by-one.

\subsection{Performance Analysis}

A performance analysis of the simulation was conducted. This was done in order to study the efficiency of both the simulation as a whole and the underlying multi-hop partitioning algorithm (as implemented in JavaScript) as the number of vehicles increased within a confined space. The result was a greater understanding of some of the application's performance limitations. The multi-hop partitioning algorithm was identified as taking the bulk of the computation time. Future performance analysis should study the effects of keeping vehicle density constant as vehicle count changes (see below).

\begin{figure*}[!htb]
\centering
\includegraphics[width=6.5in]{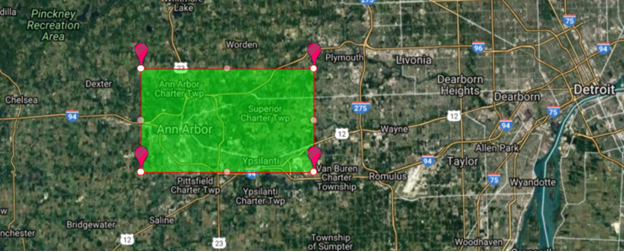}
\caption{Geographical location of the sample dataset used for simulation}
\label{rectangle}
\end{figure*}

\subsubsection{Process}

The performance testing was done by artificially generating BSM data files and measuring the simulation's performance in milliseconds when run on these files. Each generated file had a different number of vehicles per timestamp ($N$); some had as few as 1 vehicle and others had as many as 200 vehicles per timestamp. All vehicles' positions were constrained within a fixed rectangular area of 348.16 km² in Ann Arbor, Michigan (where the Safety Pilot study took place). The vehicles would appear at random points within the green rectangle in Figure \ref{rectangle}. Rather than only appearing on road surfaces, to simplify the generation process they could appear at random anywhere within this rectangle. Also, their positions in one timestamp did not influence their positions in the next timestamp. In the generated data, vehicles were confined to this given territory. Thus, as the number of vehicles was increased their density (in vehicles per kilometer) rose quickly. 

For test files that contained many vehicles, fewer timestamps were included. This was done to normalize the test file sizes to less than 5000 KB. It was observed during the experiment that the application incurred a significant amount of delay when files larger than about 4500 KB were uploaded.

Note: the GPS coordinates of the confining rectangle for generated data are as follows, as seen in Figure \ref{rectangle}: (42.356186, -83.522030), (42.356186, -83.816270), (42.226673, -83.522030), (42.226673, -83.816270).

\begin{figure*}[!htb]
\centering
\includegraphics[width=5.8in]{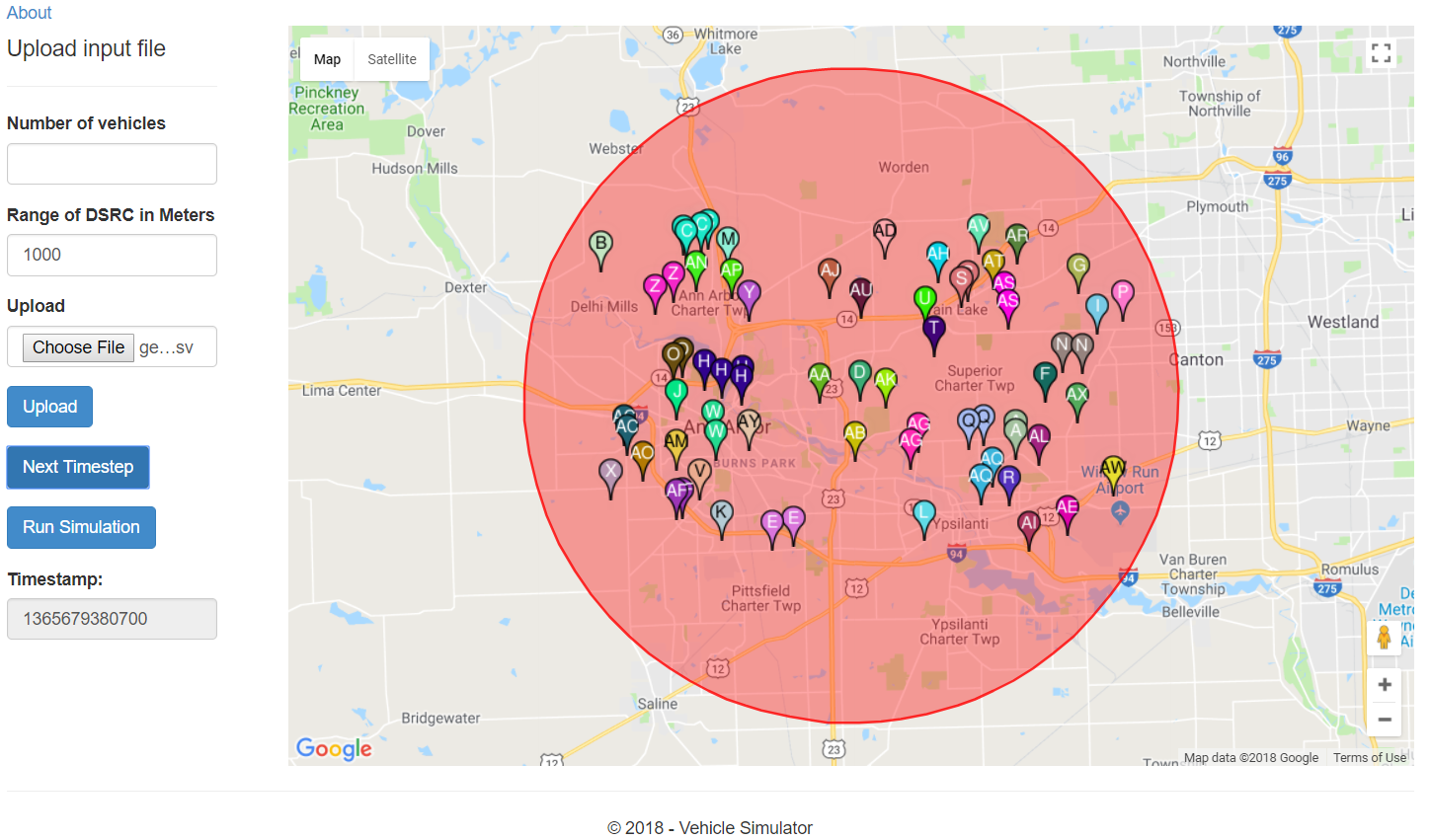}
\caption{Graphical interface of the simulator with sparse distribution of nodes}
\label{low}
\end{figure*}

\begin{figure*}[!htb]
\centering
\includegraphics[width=5.8in]{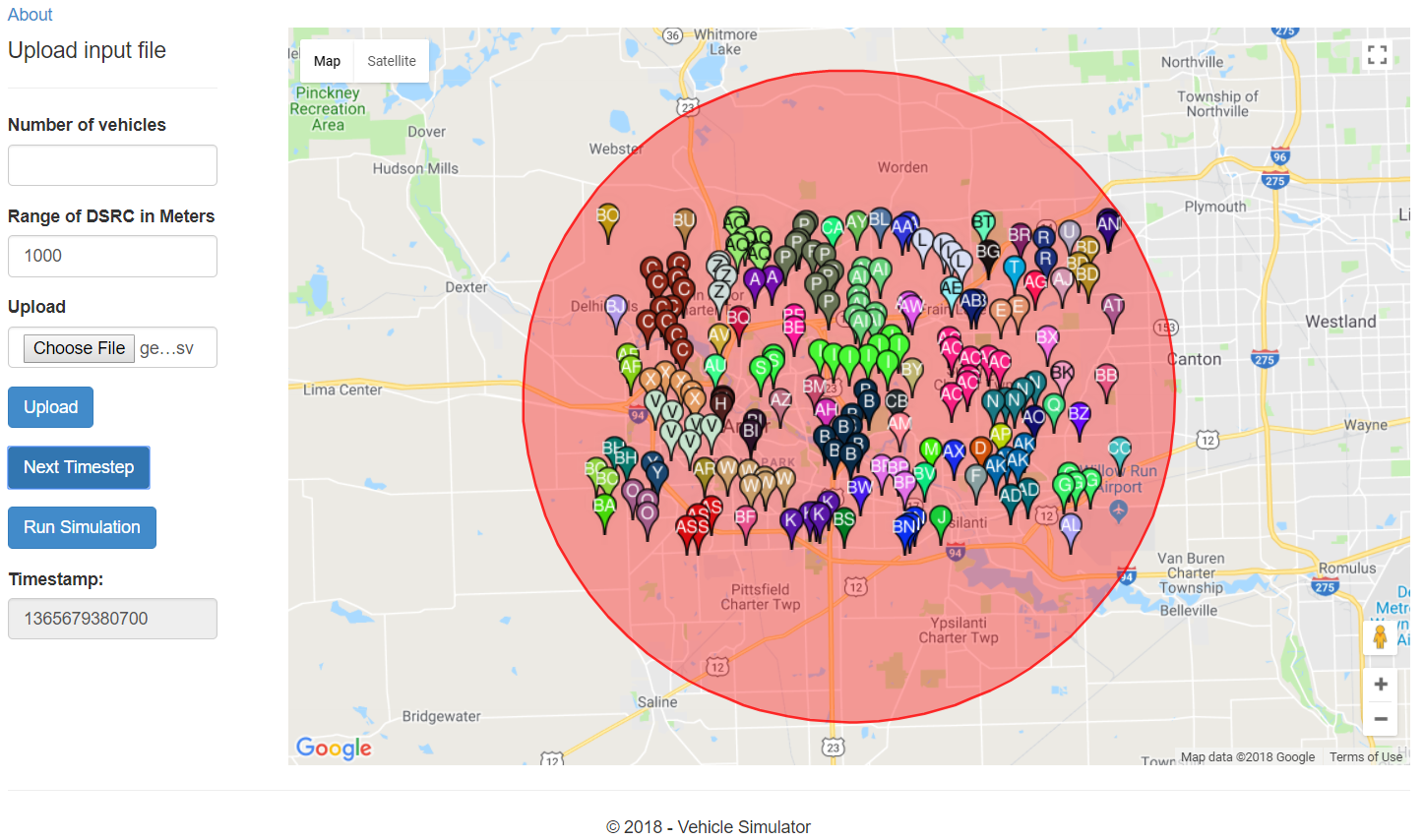}
\caption{Graphical interface of the simulator with high density of nodes}
\label{high}
\end{figure*}

From Figure \ref{low}, one can see the graphical interface for the simulator where few vehicles are sparsely distributed on the area. In Figure \ref{high}, we see a dense distribution of vehicles and their corresponding partitions. In Figure \ref{high}, the time to calculate the connectivity matrix has become very noticeable. The number of partitions has increased to its peak and will begin to lower if additional vehicles are introduced (see Figure \ref{partitionsTrend}).

\subsubsection{Results and Conclusions}

\begin{figure}[!htb]
\centering
\includegraphics[width=0.5\textwidth]{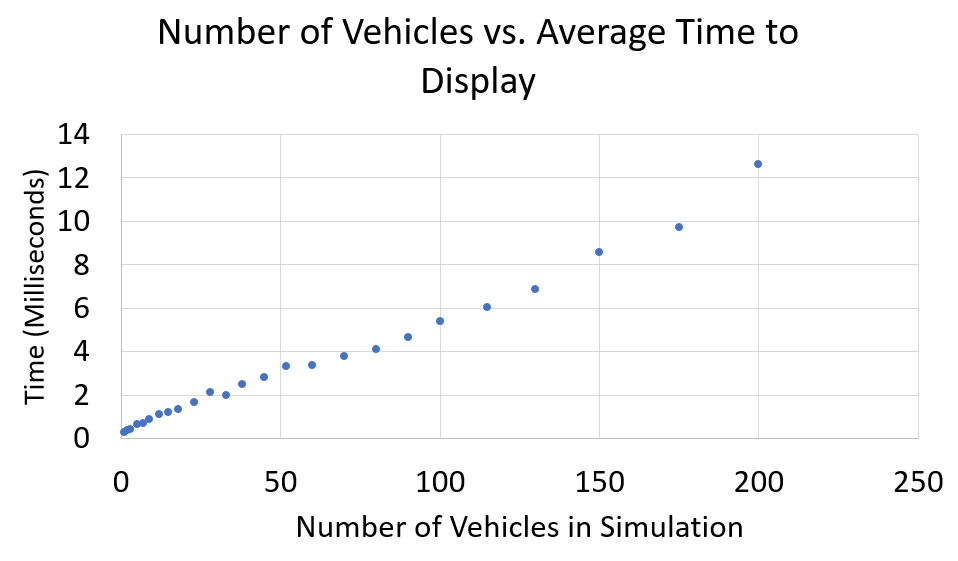}
\caption{Average time to render graphic display}
\label{displayTrend}
\end{figure}

\begin{figure}
\centering
\includegraphics[width=0.5\textwidth]{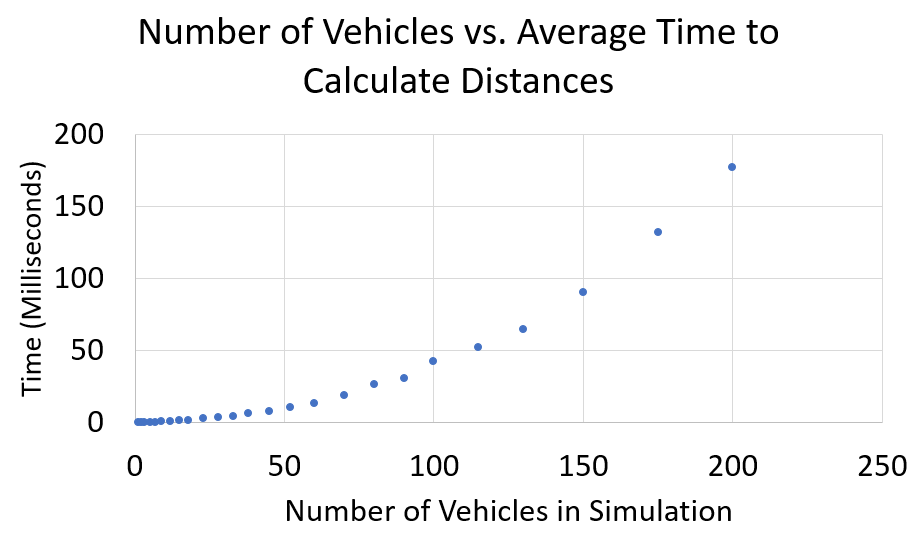}
\caption{Average time to calculate distances between vehicles}
\label{distanceTrend}
\end{figure}

\begin{figure}
\centering
\includegraphics[width=0.5\textwidth]{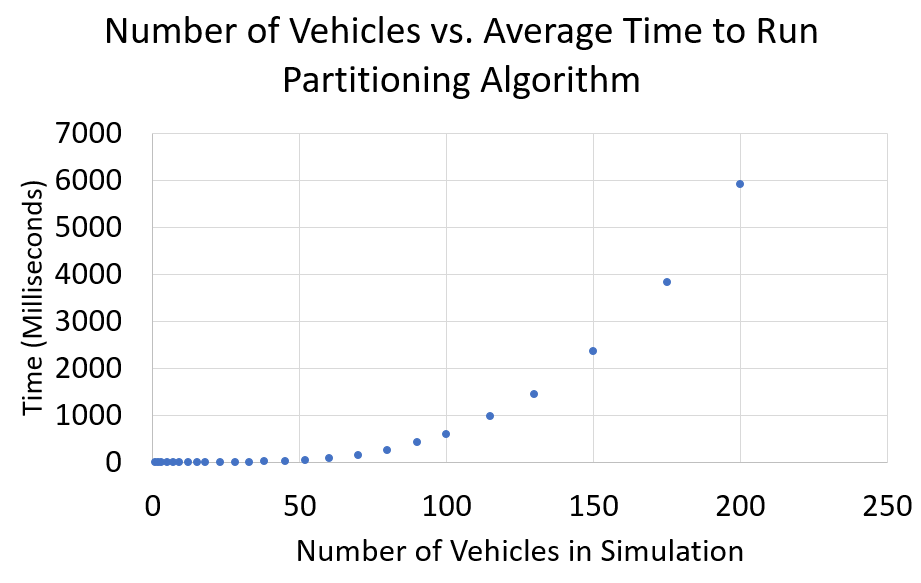}
\caption{Time complexity of partitioning algorithm}
\label{multihopTrend}
\end{figure}

\begin{figure}
\centering
\includegraphics[width=0.5\textwidth]{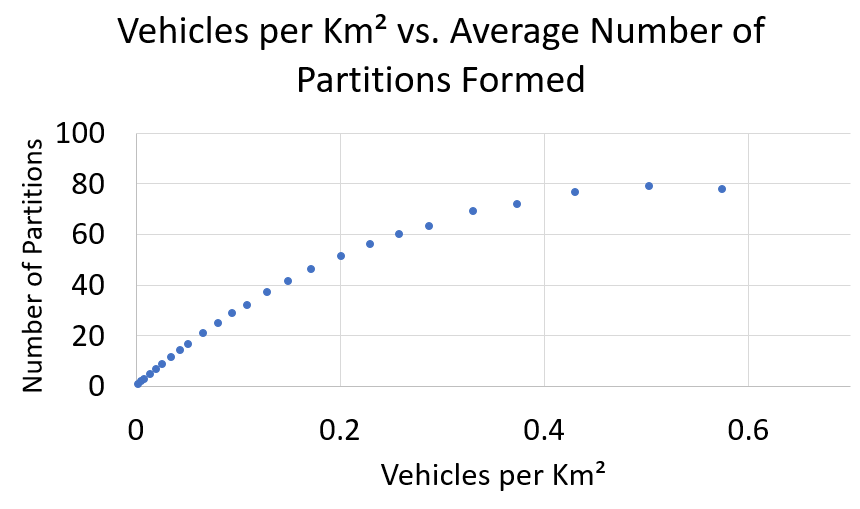}
\caption{Average number of partitions}
\label{partitionsTrend}
\end{figure}

The completion time was measured in milliseconds and logged for various parts of the simulation process. Also logged was the number of individual input vehicles within the BSM timestamp. After numerous runs with different input CSVs, the log files were combed for the data and patterns were observed; see Figures \ref{displayTrend}, \ref{distanceTrend}, \ref{multihopTrend}, and \ref{partitionsTrend}.

The time needed to populate the display with pins was insignificant, even when the population of vehicles was large (see Figure \ref{displayTrend}). This was important to verify because it was necessary to determine whether delays were being caused by the internal multi-hop process or by the display medium. In Figure \ref{displayTrend}, it appears that the time required for rendering the graphic display will increase linearly as the number of vehicles increases. The display of each additional vehicle requires only the fetching of the appropriate pin image from Google's server and the display of that pin. However, the browser's caching of these pin images likely contributed to this high performance. Ultimately, time to display was insignificant (relative to other work). Note however that this depends on the strength of Internet connection.

Referring to Fig. \ref{distanceTrend}, the distance calculation became more expensive as the number of vehicles increased. Recall that a prerequisite to performing the partition calculations is the calculation of the distance from every vehicle to every other vehicle. These calculations needed to take place exactly once before the partitions of any particular timestamp could be determined. Calculating these distances involved $n^2$ amount of work. The formula for this is $n(n-1)/2$, which is close to $n^2$. The results were in accordance with this expectation. Though the calculation of distances would eventually become expensive as the number of vehicles increased, it would still be a tiny fraction of the amount of work needed to perform the partition calculations (assuming that the vehicle density was allowed to increase).

Recall that a series of consecutive matrix multiplications forms the multi-hop partitions. From Fig. \ref{multihopTrend}, we see that as the number of vehicles increased, the multi-hop algorithm became by far the most costly part of the simulation. At 200 vehicles (0.5 vehicles per km$^2$ ), it was taking 6 seconds on average to calculate the connectivity matrix of any given timestamp.

Because of the multi-hop algorithm, the simulation began to take considerable time at a certain point (see Figure \ref{multihopTrend}). This may present a problem for those that wish to use the current tool to study very large vehicle populations (with high density in excess of 200 vehicles). In the future, an `export simulation' functionality could be implemented that could allow the user to run the simulation in the background and output the results to a file, to be displayed later rather than sequentially as they are calculated.

The simulation was run each time assuming that each vehicle had a DSRC range of 1000 meters (1 km). As the number of vehicles within the rectangle increased, so did the number of partitions (Fig. \ref{partitionsTrend})--up to a saturation point. That is, the rate of increase declined as the density grew. It was observed that at a density of 0.5 vehicles per km$^2$, the number of partitions began to decrease. If even greater densities had been tested, the number of partitions would have fallen to 1. It is expected that at a density of 1 vehicle per km$^2$ (where the square root of the inverse of the density equals the transmission range) there would have been 1 partition on average. However, in real life the saturation point may be lower. Recall that the artificial data allowed vehicles to disperse themselves randomly. In real data, vehicles would limit themselves to the roads, effectively reducing the space between vehicles. Thus for real data the density at which there would have been only a single partition would have been considerably less than that of this artificial data. This is especially true when the roads are less like a grid and more like a highway, or when the DSRC transmission range is increased by flat land and few obstacles.

\subsubsection{Limitations}

The program written to artificially generate BSM data permitted vehicles to be randomly generated anywhere in the rectangular area. In reality, vehicles stay on roads and often share the same road. They do not have the freedom of going everywhere. Thus, in reality, vehicles will be less sparse and less dispersed than they are in the artificially-generated data. This means that, if real BSM data had been used for performance analysis, there likely would have been more or less partitions than were observed. The number of partitions could have thereby had an impact on the time to calculate the partitions. There is therefore a certain degree of uncertainty regarding the extent to which the simulation will perform similarly for real data. Since this application is intended to assist transportation researchers and analysts, who generally focus on real BSM data, in the future we will run the performance analyses with real data.

\subsection{Conclusion}
An intuitive visual simulator has been developed to assist transportation researchers and analysts to study BSM data. This section has documented the application and shared the details of its performance analysis. The simulation successfully generates time-ordered displays that represent vehicle positions and connectivity statuses. The complexity of the partitioning algorithm has been identified as the application's bottleneck when the number of vehicles increases within a confined space. In a future build of the tests, it will be a goal to compare the effects of vehicle population size with those of vehicle population density. For example, keeping vehicle density constant while increasing vehicle population may produce interesting results. Another advancement would be to conduct the performance analyses with real data instead of generated data using hardware-in-the-loop simulation \cite{hoque2019parallel}.

\section{Application--2: Traffic Intersection Viewer}

\IEEEpeerreviewmaketitle

\subsection{Introduction}

Today, roadway intersections are generally less efficient and less safe than other parts of the roadway. More than 50\% of crash-related fatalities and injuries in the United States occur at or near an intersection \cite{DOT}. Sudden decelerations, idle time, and accelerations from stop lead to an increased rate of fuel consumption at these locations, thereby releasing a larger quantity of environmentally-harmful emissions and incurring higher fuel expenses. Such elevations in fuel consumption and risk of collision escalate various costs to society and the environment while endangering drivers and inducing anxiety.

These inefficient and dangerous conditions are created, in part, by drivers' imperfect knowledge regarding the timing of future signal changes. If given foreknowledge of upcoming state changes, drivers would have a longer reaction time, allowing them to adjust their speed appropriately and/or gradually. Vehicle-to-infrastructure (V2I) communication provides an avenue for providing such foresight.

A traffic controller is a device that controls the traffic lights at an intersection. Traffic controllers communicate with Signal Phase and Timing (SPaT) packets. These SPaT packets contain information about the current state of the intersection as well as the timing for and the configuration of the next upcoming state.

To achieve V2I communication between a traffic controller and an automobile/driver--and to thereby deliver foresight about upcoming state changes--it is necessary to capture these SPaT packets. Then, the SPaT packets must be parsed and translated into a readily human-readable format.

This section of the technical report is focused on two similar desktop applications, which will here be referred to as Intersection Viewer (abbreviated IV) Version 1 and Intersection Viewer Version 2. IV v1 (developed by Nick Hodge and J.T. Blevins, with maintenance by Noah Carter) parses SPaT data and displays it in an intuitive graphical user interface. IV v2 (developed by Jacob Hoyos, Matthew Dale, and Noah Carter) goes a step further by not only displaying the GUI but also acting as a server and relaying the SPaT data to a single specified client machine over a local network. In IV v2, said client then displays the GUI based on the received SPaT data.

The intended use of these applications was to serve as a learning experience and a prototype for the application described in section 3 of this report, IV v3. In IV v3 (itself yet another prototype), the client-server relationship was transitioned to the cloud, and made accessible from all personal devices over LTE (and the Internet generally).

\subsection{Related Works}

Many researchers have employed SPaT packets and the mobile LTE network for their assistive V2I applications. For example, Audi America utilized LTE as the V2I communication medium for a smart traffic light information system \cite{Zweck}. Audi also developed an optimal speed advisory system that could suggest to drivers the speed at which to travel in order to reach a green light at the appropriate moment. This advisory used 3G/4G communication and a backend server \cite{Xia}.

An environmental research group from the University of California Riverside developed a similar application using SPaT data and LTE. It provided drivers advance notice about the upcoming traffic signal timing and contrasted the fuel consumption of an informed driver with that of an uninformed driver \cite{zhao2017greendrive}. Yet another similar pollution-reducing application was developed with support from the USDOT. It relayed real-time SPaT information to drivers so they could make informed choices \cite{audi}. Ahmed et. al proposed an advisory system that generates live traffic signal information and a speed advisory to help the driver reach a destination intersection on time \cite{Ahmed}. Like previous works, Ahmed et. al used SPaT information for their advisory algorithm. However, instead of LTE/4G/4G, DSRC technology was employed.

Each of the above Intelligent Transportation System (ITS) applications helped to reduce the number of unexpected stops and starts that drivers needed to perform, thereby reducing emissions, fuel use, and safety hazards. Advances in connected vehicle technology create opportunities to develop many such applications that can boost drivers' awareness and thereby vehicle efficiency and safety.

Though IV v1 and IV v2, described below, are confined to desktops on a local wired network, they later became part of a system to broadcast SPaT data over the Internet--like the applications of Audi and California Riverside. This would allow for multiple off-site mobile clients to receive the data for display. It would enable a driver to quickly view the status of a traffic signal from anywhere with a mobile Internet connection. The desktop products, described here in section 2, served as both a prototype and a benchmark for the development and testing of the next IV iteration.


\subsection{Architecture and Implementation}

\subsubsection{Overall Architecture}

The system architecture of IV v2 consists of a traffic controller, a switch, a server machine, and a client machine. (See Figure \ref{system}.) All devices are connected over Ethernet and must be on the same subnet.

A Siemens m60 series traffic controller was used, which is a modern controller employed by various municipalities. Both the IP address of the traffic controller and the single IP address to which it sends SPaT data are configurable within the controller's setup menu. While running, the traffic controller constantly transmits SPaT packets ten times per second over the wire.

The server machine receives SPaT data from the traffic controller. It performs the translation from a byte stream to a C\# object that is interpreted by the server's GUI. The server then has the option to forward the SPaT to a single specified client machine ten times per second. The client builds the GUI in the same way that the server does, such that the GUI is potentially visible on both devices.

\begin{figure}
\centering
\begin{center}
\includegraphics[width=3.2in]{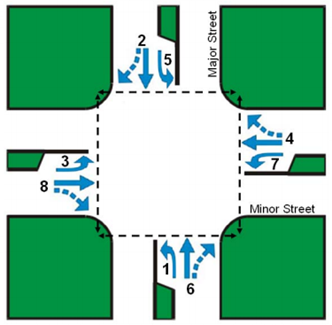}
\caption{A standard 8-phase intersection, as used for the Intersection Viewer}
\label{statuses}
\end{center}
\end{figure}

Most intersections have states which are defined by ``phases.'' Many possible phase configurations exist. IV v1-v3 are all built to interpret a regular 8-phase intersection exclusively. They cannot handle other types of intersections. Figure \ref{statuses} illustrates a classic 8-phase intersection.

\begin{figure}[h!]
  \centering
  \begin{tabular}{@{}c@{}}
    \includegraphics[width=\linewidth]{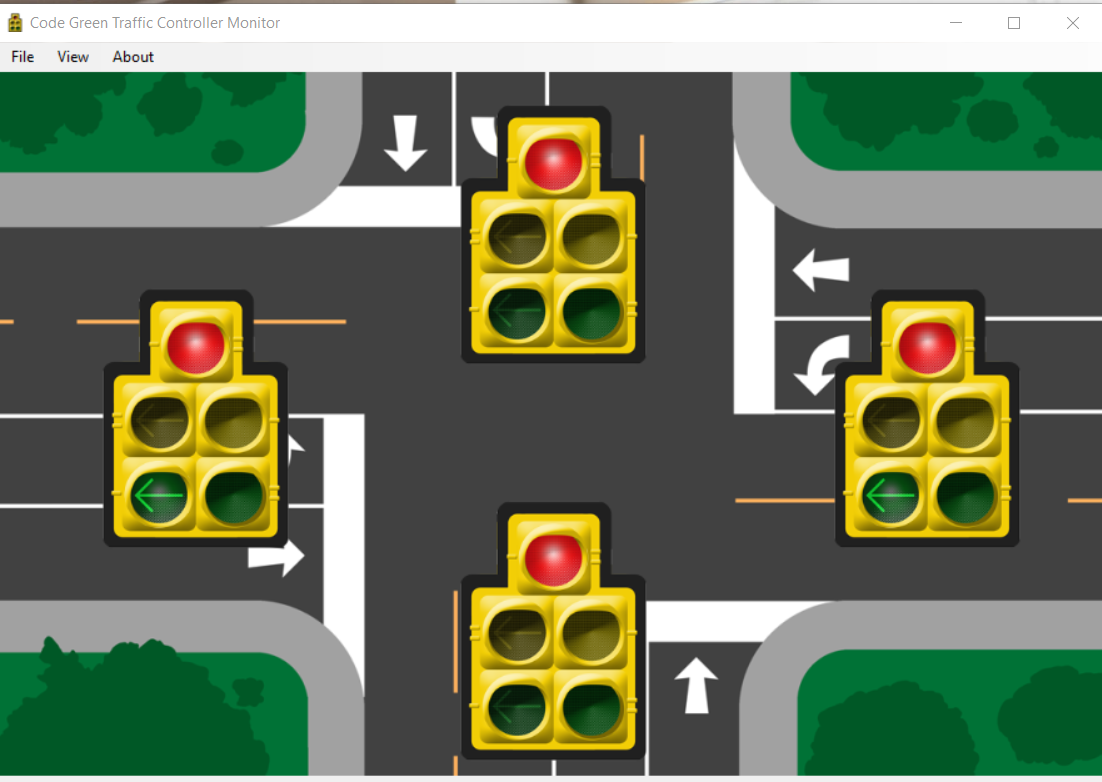} \\[\abovecaptionskip]
    \small (a) 
      \label{GUI1}
  \end{tabular}

  \begin{tabular}{@{}c@{}}
    \includegraphics[width=\linewidth]{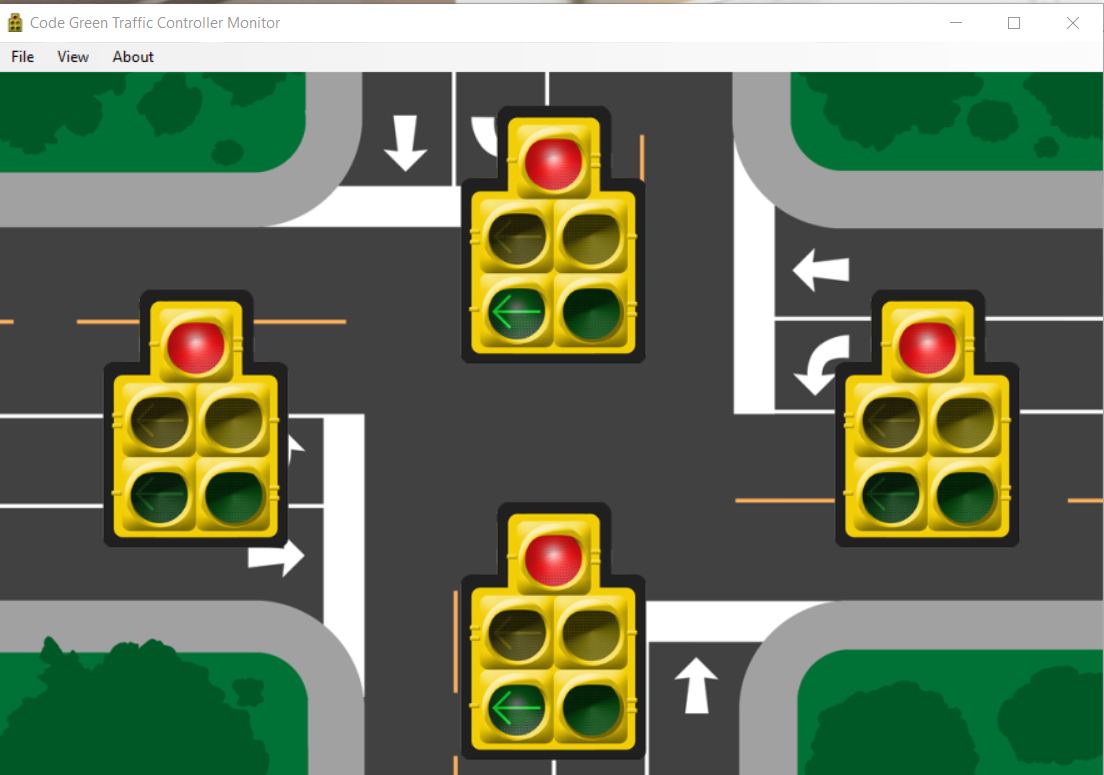} \\[\abovecaptionskip]
    \small (b) 
  \label{GUI2}
  \end{tabular}
  \begin{tabular}{@{}c@{}}
    \includegraphics[width=\linewidth]{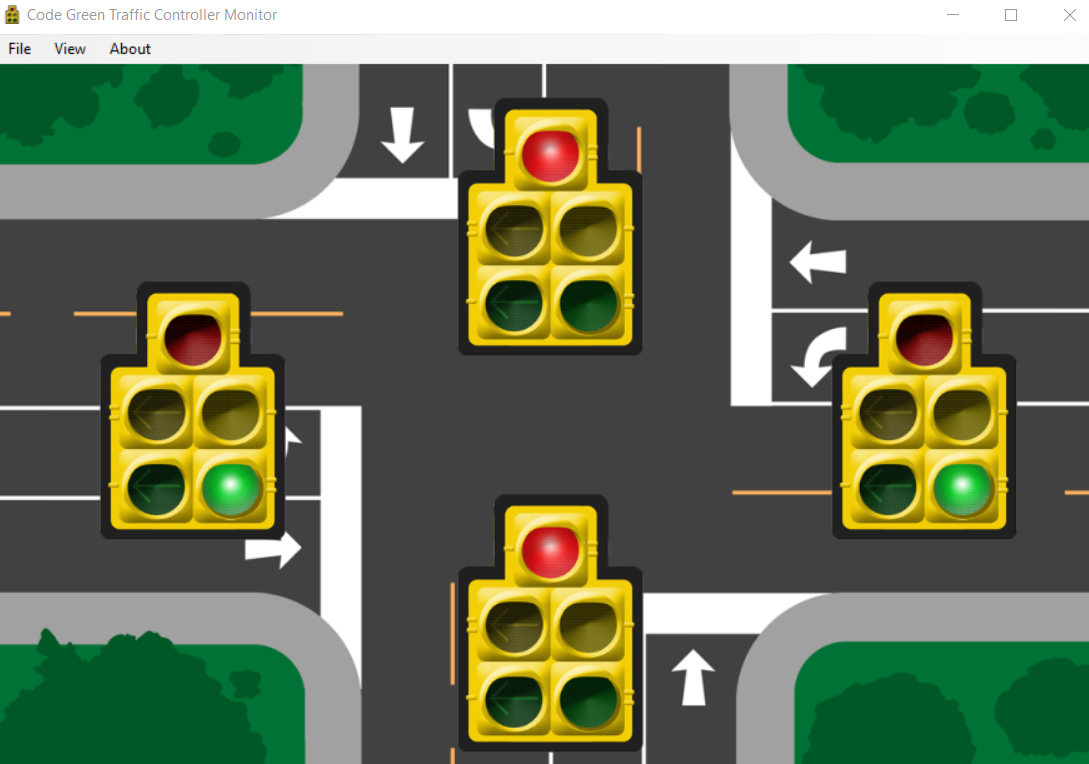} \\[\abovecaptionskip]
    \small (c) 
      \label{GUI2}
  \end{tabular}

  \caption{GUI of Intersection Viewer Version 1}\label{GUI}
\end{figure}

Figure \ref{GUI}(a), \ref{GUI}(b), and \ref{GUI}(c) demonstrate the GUI of IV v1, which was later mimicked (but modified) by IV v2. In Figure \ref{GUI}(a), green arrows are lit on the minor street to represent the left turns which correspond to phase 3 and 7.  In Figure \ref{GUI}(b), left turn lights are lit green for the major street, corresponding to phase 1 and 5. Figure \ref{GUI}(c) shows that the intersection is permitting the through movements of phase 4 and 8. During phase changes, green lights are be replaced by yellow lights.

\begin{figure*}
\centering
\includegraphics[width=6in]{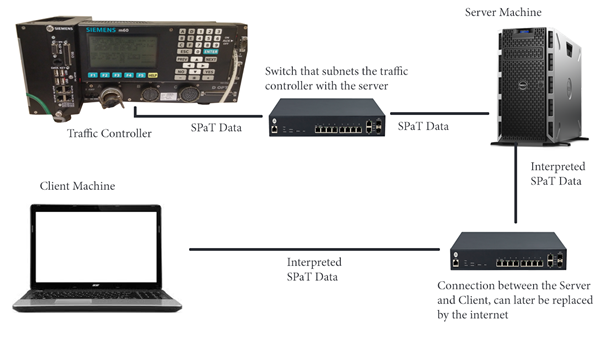}
\caption{Overall Communication Architecture of Intersection Viewer Version 2}\label{system}
\end{figure*}

\subsubsection{Notes on Implementation and Performance}

The following notes are provided primarily for the benefit of the maintenance programmer or those wishing to recreate this local client-server intersection visualization. More elaborate implementation-specific details are described in a previous article including instructions regarding the GUI and subnetting \cite{Noah2018SoutheastCon}. Note that many of the networking difficulties and shortcomings faced during the implementation of IV v2 were overcome or abstracted away by IV v3.

The server application of IV v2 utilizes the C\# implementation of the UDP Client to read in SPaT data streams. In the SPaT packets, bytes 211, 213, and 215 (representing red, yellow, and green lights respectively) are relevant to the display. In IV v2, however, byte 211 was ignored due to the assumption that all lights which do not possess a yellow or green light at a given time should be red. (This assumption is only valid in the very limited scope of a regular 8-phase intersection. It was not made by IV v3.) Also note that while a SPaT object is defined as containing 245 bytes, the last four bytes are only appended in traffic intersections which contain pedestrian call buttons.

The active traffic lights as determined from the SPaT are mapped to static images of lights that appear and disappear based on the activation of the appropriate bits. That is, they change when the server detects that the traffic controller has changed its output on the relevant bytes. The check for such change is performed on a regular interval of ten times per second.

There is a temporary delay before the GUI is in sync with the actual traffic signals. This initial lag is even more pronounced on the client. However, once 30 seconds or so passes, the traffic controller and both GUIs should become permanently synchronized. It was suggested that this initial delay might accumulate with each individual connection to the server, had multiple clients been a possibility for IV v2. (See the maintenance paper for more information.)

The server of IV v2 transmits the SPaT in a similar fashion to the way in which the traffic controller transmits the data; that is, the server sends the original byte stream out a specified port to a pre-defined IP address at a rate of once per decisecond. This means that the server and client must each know the other's IP addresses and use a common port number. Additionally, whereas the server requires a known IP, the client must connect before communication can begin. The client, in turn, mimics the way in which the server receives and interprets SPaT data.

The problems, limitations, and complications outlined above were ameliorated by the next iteration of Intersection Viewer, developed by Noah Carter and described in the following section.

\section{Intersection Viewer, Version 3}

\IEEEpeerreviewmaketitle

\subsection{Section Introduction}

Both Version 2 and Version 3 of the Intersection Viewer are designed to allow users to monitor an intersection's status by accessing a central server. In version 3, the server that clients access was changed from a machine on the local network to a public IP in the cloud. The client's desktop application was replaced by a web app.

IV v3 allows users to remotely monitor the status of a roadway intersection using the browser of a personal device, such as a laptop or mobile phone. As a web app, this implementation can be used on the web browser of any device without the need to install additional software. Simple asynchronous HTTP requests are made to send and fetch the data. Whereas IV v2 was limited to a single client with a known IP address, in IV v3 any number of clients and any platform which utilizes HTTP can access the data structures used to represent the SPaT data.


\subsection{System}
\subsubsection{Architecture and Data Flow}

The National Transportation Communications for Intelligent Transportation System Protocol (NTCIP) is a common protocol used by many transportation devices. Just as with IV v2, the third iteration of IV is designed to interpret SPaT data sent over UDP in NTCIP. Testing was, again, done with a Siemens m60 series traffic controller. This was the same type of controller used by the local Johnson City, Tennessee Traffic Division.

The following describes the data flow in IV v3. (See also Figure \ref{system}.) The traffic controller continuously transmits live SPaT data in UDP over Ethernet. A laptop (or other server-like device) is connected to the same local network. The laptop receives and parses the SPaT data into a data structure which can be understood by the client. The human-readable parsed data and the visual GUI representation originally designed in IV v1 are displayed on the laptop. Meanwhile, upon receiving a changed SPaT configuration the laptop transmits the client-bound data structure to a server in Oregon via an HTTP request. The server, which is hosted by Amazon Web Services (AWS), contains an ASP.NET controller which receives the updated data structure and saves it to a file (via an overwrite) on the AWS server. Next, various client machines (such as phones, tablets, or desktops) navigate to a public URL (also hosted by AWS) in their web browsers. The returned HTML page contains an Ajax call that repeatedly requests the saved, client-readable SPaT data structure from the AWS server. Upon receiving the data file, the client's page will, if necessary, use JavaScript to interpret the returned data structure and update a visual display of the intersection. That visual display represents the status of the 8-phase intersection.

\begin{figure*}
\centering
\includegraphics[width=6in]{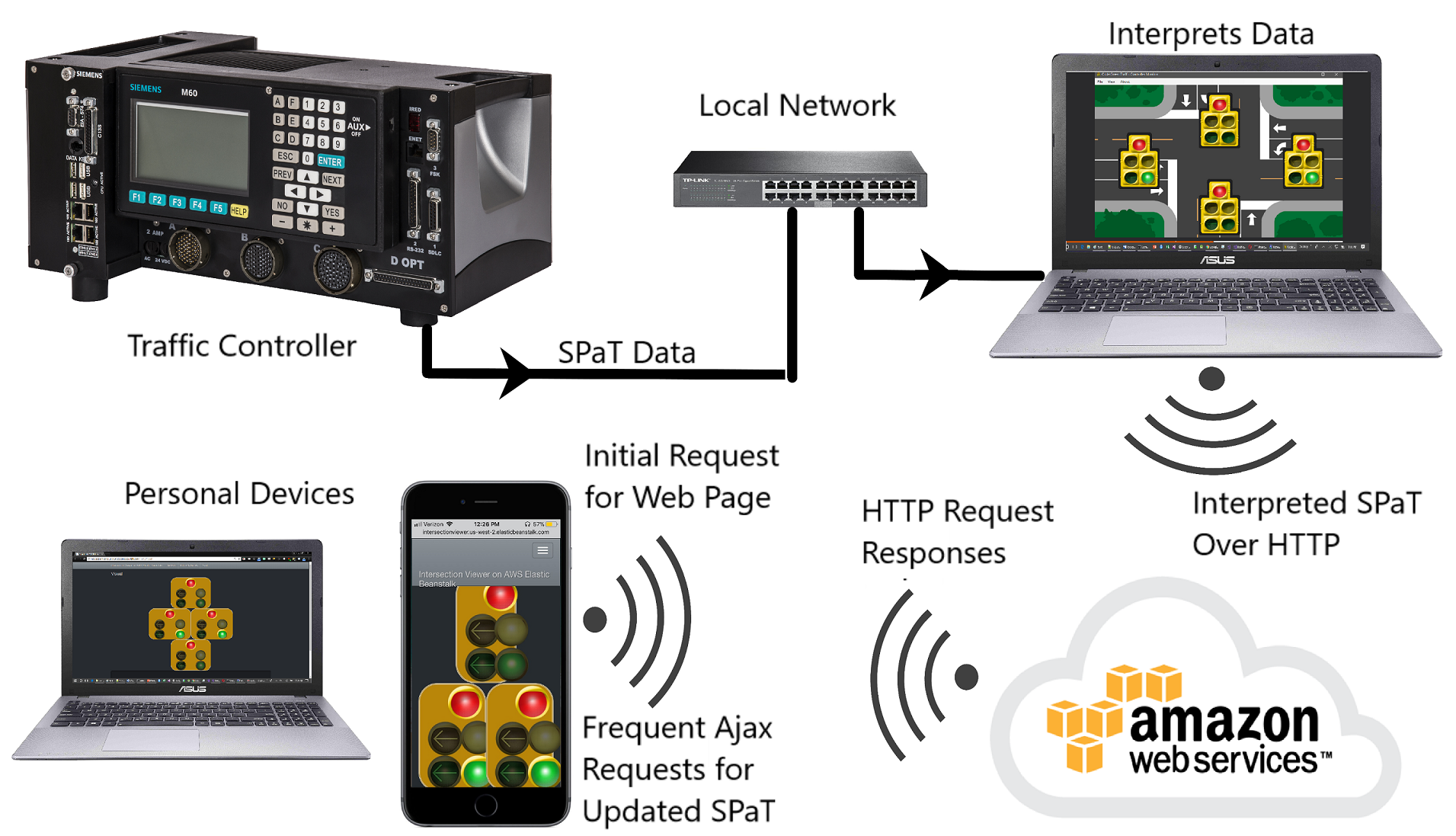}
\caption{Overall Communication Architecture for Intersection Viewer Version 3}\label{system}
\end{figure*}

\subsubsection{Further Comparison to Version 2}

The goal of IV Version 2 was to receive information from the traffic controller on an external device, including clients not directly connected to the traffic controller itself. The goal of IV Version 3 was to expand the definition of ``client" to include any devices with a web browser. The client machines need not be connected in any way with the local network of the server machine. They need to access AWS cloud resources only. Under this architecture, the subnet complications were overcome.

The details related to connection between the traffic controller and the local laptop/``server" did not change. For example, just as with IV v2, in IV v3 the traffic controller must be configured to submit data to the  IP address of the server. Refer to the server documentation in section 2.

Just as in Version 2, in IV v3 the local server (laptop) can run without a destination to which it egresses SPaT data. Also, it can still transfer the data over HTTP to a local machine running the appropriate ASP.NET environment. However, by default it is configured to send data to an existing remote, public ASP.NET controller. (Obviously, such remote calls require an Internet connection.)

The server was configured to transfer the data to AWS at a higher frequency than the previous egress frequency. That is, it no longer pauses for 0.1 second before transmitting. Instead, whenever a new SPaT configuration is detected the server makes an asynchronous HTTP call and waits for a response. It was noted that if a significant delay was allowed, and if HTTP calls were made too often, the server would gradually fall farther and farther behind the traffic controller, eventually resulting in all remote clients displaying inaccurate views. (Removing this delay from the server also decreased the initial synchronization time.)

The clients, similarly, have a very small delay. Their JavaScript waits for only a single millisecond before making a new request to AWS for updates. 

\subsection{Testing and Latency}

\subsubsection{General Observations}

After several hours of uninterrupted testing, the cell phone used was observed to have a high temperature. It may be beneficial to observe the extent to which the client's resources are used by the Ajax calls, which currently pull data from the server at a rate of once per millisecond. It is probable that a lower update frequency would reduce the cost of the software while not introducing noticeable latency.

Testing during March 2018 involved 16 hours of cumulative program runtime; as a result, the AWS server instance reached 85\% of the monthly free tier limit for file inputs and outputs, with 8.5 million I/Os. That is, the server read or wrote to a file 8.5 million times. This rate is not sustainable. Writing to a file has the advantage that the file is easily available to all HTTP clients, but opening, writing to, reading from, and closing that file is expensive (computationally and financially). It is suggested that another approach be taken to allow all clients to obtain the stored information--without sacrificing the current low latency (see below). TCP sockets were suggested as one such alternative to maintaining a file.

\subsubsection{Testing with a Live Intersection}

Once ready, the IV v3 application was tested at the Johnson City, Tennessee Traffic Division. The traffic engineers directed the traffic controller for the intersection of Indian Ridge Rd and West Market St to send its SPaT data to the IP address of the researchers' laptop. This intersection was largely a classic 8-phase intersection, with W Market being the major and Indian Ridge being the minor (see below for qualification). The laptop received the SPaT data and, using the previously discussed method, sent it to the AWS server. Multiple cell phones then accessed the data over LTE, displaying the intersection's state live.

Surprisingly, the delay between the laptop's display and the cell phones' displays was, to the human eye, observable but not significant. It was estimated at less than 0.2 seconds. This meant that the data was traveling from the laptop to the AWS server instance in Oregon, being requested by the phones, and being transmitted to and displayed by the phones with, from a human visual perspective, practically no latency.

As a way of comparison, the actual intersection was then observed with two methods: via the division's live-feed traffic video cameras and in-person. The video received by the division’s traffic cameras was observed to have a noticeably greater latency than that of the research application. On the phones, the display would update roughly 1 second sooner than the traffic division's live videos. (This may have been because of the way the cameras were configured or because the cameras needed to transport large video files while IV v3 needed only to make HTTP requests with small content loads.) When physically standing outside at the intersection, the delay between the actual traffic lights and the phones were, likewise, hardly noticeable to the human eye. The delay was approximately 0.3 seconds (as repeatedly measured with a camera and a stopwatch).

For the most part, the display was accurate, in addition to being fast. However, there were unexpected inaccuracies in the display. For example, a light would show as green on the phone when in reality both green and green left-turn were triggered. This was due, in part, to the fact that the particular intersection was using unanticipated phase logic. Though 8-phase, the intersection utilized "overlaps," in which the activation of a single light can be triggered by the activation of more than one individual phase. Removing this display bug from the research application will simply be a matter of modifying the way the SPaT data structure is interpreted.

\subsection{Research Benefits}

Intersection Viewer Version 3 is not intended as a finished product, but rather a proof of concept and prototype for further development. A team of ETSU graduate students is currently developing an app that should give drivers foreknowledge about the upcoming status of intersections--similar to those developed in by California Riverside and others. Part of their project requires a central server to relay SPaT data to mobile devices over the wireless mobile network. AWS was discussed as the potential host for the centralized portion of their application, but there were concerns about latency. (Minimizing latency is critical to delivering timely notifications to drivers.) The low latency of IV v3 has now demonstrated that AWS and the mobile network are suitable communication mediums for their application. Hence, the researchers at East Tennessee State University's Vehicular Networking Lab plan to use IV Version 3 as a starting point for the AWS-hosted, centralized portion of their safety and efficiency V2I application.

\subsection{Upcoming Modifications}

Currently, the application parses SPaT data from the controller into an arbitrary, homegrown data structure; the client machines are set up to interpret that format. SAE J7235 is a new standardized format that has been recently developed by the Society of Automotive Engineers to make SPaT communications uniform. A new task is to update the message format of the IV v3 application to the new standard message format for SPaT data, SAE J7235.

Also, the application presently sends only the light data, which contains the light statuses of the intersection. Eventually, the application should also send the ``map data," which communicates other details about the intersection such as its location.

IV v1-v3 are all limited in that they can only interpret the SPaT data from eight-phase intersection configurations. The graduate students will likely want to remove this limitation by giving the application the ability to display other types of intersections. Lastly, the CSS of the client page could be enhanced to make the HTML display more user-friendly on mobile devices. See Figure \ref{system} for the current view on a mobile device.

\subsection{Conclusion}
Intersection Viewer Version 3 provides a working example architecture and foundation for other applications which require a centralized server to host SPaT data. It can be modified and improved to assist with the development of centralized V2I applications.


\end{document}